\documentclass[12pt]{article}
 \usepackage{epsfig}
 \def\be{\begin{equation}}
 \def\ee{\end{equation}}
 \def\bea{\begin{eqnarray}}
 \def\eea{\end{eqnarray}}
 \usepackage{graphicx}

 \catcode`\@=11
 \def\lsim{\mathrel{\mathpalette\@versim<}}
 \def\gsim{\mathrel{\mathpalette\@versim>}}
 \def\@versim#1#2{\vcenter{\offinterlineskip
 \ialign{$\m@th#1\hfil##\hfil$\crcr#2\crcr\sim\crcr } }}
 \catcode`\@=12

 \catcode`@=12
\begin{document}
 \thispagestyle{empty}
 \begin{flushright}
 UCRHEP-T611\\
 Mar 2021\
 \end{flushright}
 \vspace{1.0in}
 \begin{center}
 {\LARGE \bf Naturally Light Dirac Neutrinos from $SU(6)$\\}
 \vspace{1.2in}
 {\bf Ernest Ma\\}
 \vspace{0.2in}
{\sl Department of Physics and Astronomy,\\ 
University of California, Riverside, California 92521, USA\\}
\end{center}
 \vspace{1.8in}

\begin{abstract}\
A known mechanism for obtaining naturally light Dirac neutrinos is 
implemented in the context of $SU(6) \to SU(5) \times U(1)_N$.  
\end{abstract}

 \newpage
 \baselineskip 24pt

\noindent \underline{\it Introduction}~:~
Whether or not neutrinos are Majorana or Dirac is a fundamental issue in 
particle physics.  Experimentally, there is still no evidence for one or 
the other, although it is known that at least two neutrinos must have masses, 
because of neutrino oscillations~\cite{pdg20}.  Theoretically, the 
standard model (SM) requires only neutrinos in the left-handed 
$SU(2)_L \times U(1)_Y$ doublets $(\nu,l)_L$.  The singlet $\nu_R$ 
is not necessary because it is trivial under the SM gauge group 
$SU(3)_C \times SU(2)_L \times U(1)_Y$.  To have a Dirac neutrino, $\nu_R$ 
must exist.  To justify its existence, a gauge extension of the SM is 
often considered, either $U(1)_Y \to U(1)_Y \times U(1)_{B-L}$~\cite{d79} 
or $U(1)_Y \to SU(2)_R \times U(1)_{B-L}$~\cite{mm80}, which may be 
incorporated into the grand unified structure of $SO(10)$.  The breaking 
of $U(1)_{B-L}$ [and $SU(2)_R$] is usually assumed without hesitation 
to allow $\nu_R$ to obtain a large Majorana mass, so that $\nu_L$ gets 
a tiny seesaw Majorana mass, as is well-known.

There is actually another option.  This breaking does not have to be 
$\Delta L = 2$.  If it is $\Delta L = 3$ for example, then neutrinos 
are Dirac.  This was first pointed out~\cite{mpr13} for a general $U(1)_X$ 
symmetry and applied~\cite{ms15} to $U(1)_L$ for Dirac neutrinos.
However, this mechanism does not by itself explain why the neutrino 
Higgs Yukawa couplings are so small.

To overcome this problem, the mechanism of Ref.~\cite{m01} is the simplest 
solution.  The idea is to have at least two Higgs doublets, say 
$\Phi = (\phi^+,\phi^0)$ and $\eta = (\eta^+,\eta^0)$ which are 
distinguished by some symmetry, so that $\bar{\nu}_R \nu_L$ couples to 
$\eta^0$, but not $\phi^0$.  This symmetry is then broken by the 
$\mu^2 \Phi^\dagger \eta$ term, under the condition that $m^2_\Phi < 0$ 
and $m^2_\eta >0$ and large.  In that case, the vacuum expectation value 
$\langle \eta^0 \rangle$ is given by $-\mu^2 \langle \phi^0 \rangle/m^2_\eta$, 
which is naturally small, implying thus a very small Dirac neutrino mass. 
In the original application~\cite{m01}, $\nu_R$ is also allowed a large 
Majorana mass, hence the mass of $\nu_L$ is doubly suppressed.  In that 
case, $m_\eta$ could well be of order 1 TeV.  On the other hand, if the 
symmetry and the particle content are such that $\nu_R$ is prevented 
from having a Majorana mass, then a much larger $m_\eta$ works just as 
well for a tiny Dirac neutrino mass.

Recently this idea has been applied~\cite{m21,m21-1} using a gauge $U(1)_D$ 
symmetry under which the SM particles do not transform, but $\nu_R$ and 
other fermion singlets do.  The $U(1)_D$ symmetry is broken by singlet 
scalars which transform only under $U(1)_D$.  The bridge between the SM 
and this new sector is a set of Higgs doublets which transform under both. 
The particle content is chosen such that global lepton number is conserved 
as well as a dark parity or dark number.

In this paper, instead of the {\it ad hoc} $U(1)_D$ symmetry, $\nu_R$ is 
identified as part of the fundamental representation of $SU(6)$ which 
breaks to $SU(5) \times U(1)_N$.  Following a recent analysis~\cite{m21-2}, 
it is shown how naturally small Dirac neutrino masses occur in this 
context.

\noindent \underline{\it Description of Model}~:~
Starting with the well-known $SU(5)$ model~\cite{gg74} of grand 
unification, an extension to $SU(6)$ is straightforward~\cite{ikn77,y78}. 
Instead of having the anomaly-free combination of $5^*$ and $10$ under 
$SU(5)$ for each family, there should be now two $6^* = (5^*,-1) + (1,5)$ 
and one $15 = (10,2) + (5,-4)$ under $SU(6) \to SU(5) \times U(1)_N$. Let
\begin{equation}
6^*_{F1}  = \pmatrix {d^c \cr d^c \cr d^c \cr e \cr \nu \cr N}, ~~~ 
6^*_{F2}  = \pmatrix {D^c \cr D^c \cr D^c \cr E^- \cr E^0 \cr \nu^c}, ~~~ 
15_F = \pmatrix {0 & u^c & u^c & -u & -d & -D \cr -u^c & 0 & u^c & -u & -d 
& -D \cr u^c & -u^c & 0 & -u & -d & -D \cr u & u & u & 0 & -e^c & -E^+ \cr 
d & d & d & e^c & 0 & -\bar{E}^0 \cr D & D & D & E^+ & \bar{E}^0 & 0}.
\end{equation}
Their $SU(3)_C \times SU(2)_L \times U(1)_Y \times U(1)_N$ assignments are 
listed in Table~1.  Note that all are left-handed.
\begin{table}[tbh]
\centering
\begin{tabular}{|c|c|c|c|c|c|c|}
\hline
fermion & $SU(5)$ & $SU(3)_C$ & $SU(2)_L$ & $U(1)_Y$ & $U(1)_N$ & $U(1)_X$ \\
\hline
$d^c$ & $5^*$ & $3^*$ & 1 & 1/3 & $-1$ & 1 \\ 
$(\nu,e)$ & $5^*$ & 1 & 2 & $-1/2$ & $-1$ & 1 \\ 
$N$ & 1 & 1 & 1 & 0 & $5$ & 1 \\ 
\hline
$D^c$ & $5^*$ & $3^*$ & 1 & 1/3 & $-1$ & 2 \\ 
$(E^0,E^-)$ & $5^*$ & 1 & 2 & $-1/2$ & $-1$ & 2 \\ 
$\nu^c$ & 1 & 1 & 1 & 0 & $5$ & 2 \\ 
\hline
$(u,d)$ & 10 & 3 & 2 & 1/6 & 2 & 0 \\ 
$u^c$ & 10 & $3^*$ & 1 & $-2/3$ & 2 & 0 \\
$e^c$ & 10 & 1 & 1 & 1 & 2 & 0 \\ 
\hline
$D$ & 5 & 3 & 1 & $-1/3$ & $-4$ & 0 \\ 
$(E^+,\bar{E}^0)$ & 5 & 1 & 2 & 1/2 & $-4$ & 0 \\ 
\hline
\end{tabular}
\caption{Fermion content of $SU(6) \to SU(5) \times U(1)_N$ model.}
\end{table}
The added $U(1)_X$ is a global symmetry imposed on the dimension-four 
couplings of the resulting Lagrangian, but softly broken by bilinear and 
trilinear scalar terms.  The scalars which break the $SU(6)$ symmetry and 
allow these fermions to acquire masses are listed in Table~2.
\begin{table}[tbh]
\centering
\begin{tabular}{|c|c|c|c|c|c|c|c|}
\hline
$\langle scalar \rangle$ & $SU(6)$ & $SU(5)$ & $SU(3)_C$ & $SU(2)_L$ & 
$U(1)_Y$ & $U(1)_N$ & $U(1)_X$ \\
\hline
$u_1$ & 35 & $24$ & $1$ & 1 & 0 & $0$ & 0 \\ 
$u_2$ & $6$ & 1 & 1 & $1$ & 0 & $-5$ & 2 \\ 
$u_3$ & $21^*$ & 1 & 1 & $1$ & 0 & $10$ & 2 \\ 
\hline
$v_1$ & $84$ & $5$ & 1 & 2 & $1/2$ & 1 & 1 \\ 
$v_2$ & $15^*$ & $5^*$ & 1 & 2 & $-1/2$ & $4$ & 0 \\ 
$v_3$ & $15^*$ & $5^*$ & 1 & 2 & $-1/2$ & $4$ & 3 \\ 
\hline
\end{tabular}
\caption{Scalar content of $SU(6) \to SU(5) \times U(1)_N$ model.}
\end{table}
The $35_S$ breaks $SU(6)$ at the garnd unification scale $u_1$ to 
$SU(3)_C \times SU(2)_L \times U(1)_Y \times U(1)_N$.  The $6_S$ breaks 
$U(1)_N$ at a lower scale.  Because it is charged under $U(1)_X$, its 
only allowed coupling is $6_S^* \times 6^*_{F_2} \times 15_F$.  Hence 
$D^cD$ and $E^-E^++E^0\bar{E}^0$ masses are proportional to $u_2$.  The 
$21^*_S$ also breaks $U(1)_N$.  The $21_S \times 6^*_{F1} \times 6^*_{F1}$ 
term yields Majorana masses $NN$ proportional to $u_3$.  The 
electroweak $SU(2)_L \times U(1)_Y$ symmetry is broken by three Higgs 
doublets.  The $84^*_S \times 6^*_{F1} \times 15_F$ term yields masses 
for $d^cd, ee^c, N\bar{E}^0$ which are proportional to $v_1$.  The  
$15_S \times 15_F \times 15_F$ term yields masses for $u^cu$ 
proportional to $v_2$, whereas the $15_S \times 6^*_{F1} \times 6^*_{F2}$ 
term yields masses for $\nu \nu^c, N E^0$ proportional to $v_3$.  Let 
these three Higgs doublets be named $\Phi_{1,2}$ and $\eta$ respectively. 
To obtain a small $v_3$, it is clear that $m^2_\eta$ must be positive 
and large, as discussed in the introduction.  Now both $\eta$ and $\Phi_2$ 
come from $15_S$ but are distinguished by $U(1)_X$.  Hence the soft  
term $\mu^2 \Phi^\dagger_2 \eta$ breaking $U(1)_X$ by 3 units is available 
to make the Dirac neutrino masses very small.  Note that $\nu^c$ does not 
acquire a Majorana mass because of the absence of a scalar $21^*$ with 4 units 
of charge under $U(1)_X$.  The two soft trilinear terms 
$84_S \times 6_S \times 15^*_S$ break $U(1)_X$ by 3 and 6 units, 
whereas $6_S \times 6_S \times 21^*_S$ does it by 6 units.

\noindent \underline{\it Residual Symmetries}~:~
Because of the $U(1)_X$ symmetry, Yukawa terms are restricted so that 
residual symmetries exist for the fermions of this model at the level of 
$SU(3)_C \times SU(2)_L \times U(1)_Y \times U(1)_N$.  The usual baryon 
number $B$ and lepton number $L$ are then conserved, together with a dark 
parity $Z_2^D$ which may be identified as $(-1)^{3B+L+2j}$, where $j$ 
is the intrinsic spin of the particle, as shown in Table~3.    
\begin{table}[tbh]
\centering
\begin{tabular}{|c|c|c|c|}
\hline
fermion & $B$ & $L$ & $Z_2^D$ \\
\hline
$u,d$ & 1/3 & $0$ & $+$ \\ 
$\nu,l$ & 0 & $1$ & $+$ \\
\hline 
$D$ & 1/3 & $1$ & $-$ \\ 
$N,E^0,E^-$ & 0 & $0$ & $-$ \\ 
\hline
\end{tabular}
\caption{Residual Symmetries of $SU(6) \to SU(5) \times U(1)_N$ model.}
\end{table}
  
The $3 \times 3$ mass matrix spanning $(N,E^0,\bar{E}^0)$ is of the form 
\begin{equation}
{\cal M}_{NE} = \pmatrix{f_N u_3 & f_{NE} v_3 & f'_{NE} v_1 \cr 
f_{NE} v_3 & 0 & f_E u_2 \cr f'_{NE} v_1 & f_E u_2 & 0}.
\end{equation}
The lightest mass eigenstate is possible dark matter.  However as shown 
in Ref.~\cite{m21-2}, because of its $SU(6)$ antecedent, $N$ may decay 
through a superheavy gauge boson in analogy to proton decay.  Note also 
that $v_3$ is very small.  If $u_3$ is absent, then $N$ gets a small seesaw 
mass proportional to $v_1 v_3/u_2$.  This makes it a candidate for very light 
freeze-in dark matter, as described in Ref.~\cite{m21-1}.  In such a 
scenario, the $21^*$ scalar is not required.

In the $15_S \times 6^*_{F1} \times 6^*_{F2}$ coupling, the $(5,-4)$ 
component of $15_S$ contains $\eta^0$ which couples to $\nu \nu^c$ 
as well as $N E^0$.  It also contains the scalar color triplet 
$\zeta = (3,1,-1/3,-4)$ which couples to both $d^c \nu^c$ and $N D^c$.  
Hence $D^c$ must have $L=-1$ and $N$ must have $L=0$.  Note that $\zeta$ 
has then $B=1/3$ and $L=1$, so that its dark parity, i.e. $(-1)^{3B+L+2j}$, 
is even as expected.  The dark quark $D$ decays through $\zeta$ to 
$d \nu N$.  This process is a three-body decay with two invisible 
particles and not easy to observe.

\noindent \underline{\it Model Characteristics}~:~
Below the breaking scale $u_{2,3}$ of $U(1)_N$, the particle content of 
the proposed model is that of the SM with the following changes.  The 
neutrinos are Dirac particles.  There are two Higgs doublets, one coupling 
to $\bar{u} u$ and the other to $\bar{d} d$ and $\bar{e} e$.  A third 
Higgs doublet is very heavy and not observable, but it has a tiny vacuum 
expectation value and couples to $\bar{\nu} \nu$.  There may also be 
a very light neutral Majorana fermion $N$ which is freeze-in dark matter 
in the case that $u_3=0$.

The above particles all interact with a new heavy gauge boson $Z_N$ coming 
from $U(1)_N$, according to their charges given in Tables~1 and 2.  The 
present collider limit~\cite{pdg20} of $Z_N$ is estimated to be a few TeV.
At or above the $Z_N$ mass scale, particles of the dark sector as well as 
one or more Higgs singlets should appear.  The $D$ quark may be easily 
produced at the collider and decay to the $d$ quark plus a neutrino 
and the dark $N$.  The doublet $(E^0,E^-)$ dark fermions also decay to 
$N$ plus a scalar or the $W^-$ gauge boson.  More details are in 
Ref.~\cite{m21-2}.

\noindent \underline{\it Concluding Remarks}~:~
It has been shown that a framework exists for naturally small Dirac 
neutrino masses in the context of $SU(6) \to SU(5) \times U(1)_N$, where 
the right-handed neutrino singlet $\nu^c$ is embedded as shown in Eq.~(1). 
With the implementation of a softly broken global $U(1)_X$ symmetry as 
given in Tables~1 and 2, the residual symmetries of baryon number $B$ 
and lepton number $L$ are preserved as given in Table~3, with dark parity 
identified as $Z_2^D = (-1)^{3B+L+2j}$.  The reason for a conserved lepton 
number parallels that of Ref.~\cite{mpr13}, i.e. the interplay of $U(1)_X$ and 
the chosen $SU(6)$ representations makes it impossible for $\nu^c$ to be 
a Majorana fermion.

The dark sector consists of $D$ leptoquarks 
and $L=0$ fermions $N$ and the vectorlike doublet $(E^0,E^-)$.  Whereas 
$N$ mixes with $E^0$ and $\bar{E}^0$, the former may be almost a mass 
eigensate and considered as dark matter.  It is presumably of order a 
few TeV, but if the $21^*$ scalar is removed from Table~2, then $u_3=0$ 
and Eq.~(2) yields a very small mass for $N$ which then becomes freeze-in 
dark matter.

Suppose the $U(1)_N$ breaking scale is much higher than a few TeV, then 
only the SM particles are observable, with the important difference 
that neutrinos are Dirac, and there are two Higgs doublets.  The other 
possible addition is the Majorana fermion $N$ as freeze-in dark 
matter.  If $u_3=0$, then its mass is proportional to $v_1 v_3/u_2$ as 
remarked earlier, which is perhaps too small because $v_3$ is responsible 
for the Dirac neutrino mass and $u_2$ is now assumed to be very large.  
However, the $21^*$ scalar may be retained, and $u_3 \neq 0$ rendered small 
(but not too small) by the same mechanism~\cite{m01} which makes $v_3$ small, 
i.e. of the form $-\mu_{23} u_2^2/m_3^2$ from the term 
$6_S \times 6_S \times 21^*_S$,  where $m_3$ is the mass of the scalar $21^*$.

\noindent \underline{\it Acknowledgement}~:~
This work was supported in part by the U.~S.~Department of Energy Grant 
No. DE-SC0008541.

\bibliographystyle{unsrt}

\begin{thebibliography}{99}
\bibitem{pdg20} P.A. Zyla {\it et al.} (Particle Data Group), Prog. Theor. 
Exp. Phys. {\bf 2020}, 083C01 (2020).
\bibitem{d79} A. Davidson, Phys. Rev. {\bf D20}, 776 (1979).
\bibitem{mm80} R. N. Mohapatra and R. E. Marshak, Phys. Lett. {\bf B91}, 
222 (1980).
\bibitem{mpr13} E. Ma, I. Picek, and B. Radovcic, Phys. Lett. {\bf B726}, 
744 (2013).
\bibitem{ms15} E. Ma and R. Srivastava, Phys. Lett. {\bf B741}, 217 (2015).
\bibitem{m01} E. Ma, Phys. Rev. Lett. {\bf 86}, 2502 (2001).
\bibitem{m21} E. Ma, arXiv:2101.12138 [hep-ph].
\bibitem{m21-1} E. Ma, Phys. Lett. {\bf B 815}, 136162 (2021).
\bibitem{m21-2} E. Ma, arXiv:2011.01398 [hep-ph].
\bibitem{gg74} H. Georgi and S. L. Glashow, Phys. Rev. Lett. {\bf 32}, 438 
(1974).
\bibitem{ikn77} K. Inoue, A. Kakuto, and Y. Nakano, Prog. Theor. Phys. 
{\bf 58}, 630 (1977).
\bibitem{y78} S. K. Yun, Phys. Rev. {\bf D18}, 3472 (1978).

\end{thebibliography}

\end{document}